\documentclass[12pt,preprint]{aastex}
\usepackage{amssymb}
\usepackage{amsmath}
\usepackage{graphicx}
\usepackage{lscape}

\def\ms{M$_{\odot}$}
\def\kms{km~s$^{-1}$}

\def\msyr{M$_{\odot}$yr$^{-1}$}

\def\mspcGyr{M$_{\odot}$pc$^{-2}$Gyr$^{-1}$}
\def\mspc{M$_{\odot}$pc$^{-2}$}
\def\dkp{dex~kpc$^{-1}$}

\begin{document}

\title{Origin and Evolution of the Abundance Gradient along the Milky Way Disk }

\author{J. Fu$^{1,2}$, J.L. Hou$^{1}$, J. Yin$^{1,2}$ and R.X. Chang$^{1}$ \\ [5mm]}

\affil{$^1$Key Laboratory for Research in Galaxies and Cosmology,
Shanghai Astronomical Observatory, the Chinese Academy of
Sciences, 80 Nandan Road, Shanghai, 200030, China \\
$^2$Graduate School, the Chinese Academy of Sciences, Beijing,
100039, China }

\email{fujian@shao.ac.cn; houjl@shao.ac.cn; jyin@shao.ac.cn,
crx@shao.ac.cn}

\begin{abstract}

Based on a simple model of the chemical evolution of the Milky Way
disk, we investigate the disk oxygen abundance gradient and its time
evolution. Two star formation rates (SFRs) are considered, one is
the classical Kennicutt-Schmidt law ($ \Psi = 0.25
\Sigma_{\rm{gas}}^{1.4}$, hereafter C-KS law), another is the
modified Kennicutt law ($\Psi  = \alpha \Sigma _{{\rm{gas}}}^{1.4}
\left( {V/r} \right)$, hereafter M-KS law). In both cases, the model
can produce some amount of abundance gradient, and the gradient is
steeper in the early epoch of disk evolution. However, we find that
when C-KS law is adopted, the classical chemical evolution model,
which assumes a radial dependent infall time scale, cannot produce a
sufficiently steep present-day abundance gradient. This problem
disappears if we introduce a disk formation time scale, which means
that at early times, infalling gas cools down onto the inner disk
only, while the outer disk forms later. This kind of model, however,
will predict a very steep gradient in the past. When the M-KS law is
adopted, the model can properly predict both the current abundance
gradient and its time evolution, matching recent observations from
planetary nebulae and open clusters along the Milky Way disk. Our
best model also predicts that outer disk (artificially defined as
the disk with $R_g \ge 8kpc$) has a steeper gradient than the inner
disk. The observed outer disk gradients from Cepheids, open clusters
and young stars show quite controversial results. There are also
some hints from Cepheids that the outer disk abundance gradient may
have a bimodal distribution. More data is needed in order to clarify
the outer disk gradient problem. Our model calculations show that
for an individual Milky Way-type galaxy, a better description of the
local star formation is the modified KS law.

\end{abstract}


\section{Introduction}

The radial abundance gradient of a disk galaxy is an essential
ingredient in an accurate picture of galaxy formation and evolution.
During the past twenty years, the existence of an abundance gradient
along the Milky Way disk has been well established by observations
using various tracers. An oxygen and/or iron abundance gradient of
about $-0.05 \sim -0.07$ \dkp\ was obtained by observing young stars
and HII regions (see Hou \& Chang 2001; Rudolph et al. 2006 and
references therein), planetary nebulae (Maciel et al. 2005; 2006)
and open clusters (Friel 1995, 1999; Carraro et al. 1998; Chen et
al. 2003). Two spiral neighbors of the Milky Way Galaxy in the Local
Group, M33 and M31 have provided further evidence for the existence
of disk abundance gradient. A relatively flatter gradient is
observed for M31, which is about $-0.04 \sim -0.05$ \dkp \
(Dennefeld et al. 1981; Blair et al. 1982). As for M33, the observed
value of oxygen abundance gradient diverges greatly at present.
Different observers have presented different values based on various
objects (mainly HII regions and B stars). Published data show that
the oxygen gradient varies from $-$0.02 \dkp\ to $-$0.16 \dkp
(Magrini et al. 2007a). A larger and more homogeneous sample of HII
regions that covers the whole M33 disk is needed in order to get
conclusive results about the real gradients. Such a project is
currently undergoing by a couple of groups (Rosolowsky \& Simon
2008; Magrini et al. 2007b).

Nevertheless, it is important to point out that the abundance
gradients for different elements are sometimes different due to
different nucleosynthetic history, and the exact values for the
Milky Way disk are still not very certain (Deharveng et al. 2000;
Daflon \& Cunha 2004; Carraro et al. 2007). This has prevented the
chemical evolution models from being constrained clearly.

Another question is how the abundance gradient along a Galaxy disk
evolves during the history of disk evolution. With the help of large
samples of Open Clusters (OCs) and Planetary Nebulae (PNe), it is
now possible to explore this question observationally. The estimated
ages of OCs and PNe of various types span a large fraction of the
age of the Galaxy. Observations of the abundances of those objects
across the Milky Way disk have provided some important information
on the past history of the abundance gradients. Indeed, current data
show that the gradient was steeper in the past (Carraro et al. 1998,
2007; Hou et al. 2002; Chen et al. 2003; Maciel et al. 2003, 2005,
2006; Magrini et al. 2008).

The observed disk abundance gradient and its evolution offer the
opportunity to test theories of disk chemical evolution and stellar
nucleosynthesis. Detailed knowledge about the disk abundance
gradient and its time evolution are crucial not only in our
understanding of the metal enrichment history along the disk, that
is the star formation history, but also very important in our
understanding of the metallicity relationship between high redshift
Dampled Lyman Alpha systems and local disks (Wolfe et al. 2005; Hou
et al. 2005). In the framework of a phenomenological scenario of
disk formation, several mechanisms may play roles in shaping the
abundance gradient and its evolution. One is the star formation
processes in the disk. Observations of local disk and starburst
galaxies show that the star formation rate (SFR) per unit area is
proportional to some power of the gas mass surface density
(Kennicutt 1998). This non-linear star formation law may result in
and largely influence the building of radial abundance gradients.
Another mechanism is the so-called ``inside-out'' disk formation. In
most classical chemical evolution models, it is generally assumed
that the disk forms by gas infalling from the outer halo, and that
the infall can be described by an exponential law, in which the
infall time scale is always assumed to be radially dependent
$\tau=\tau(r)$, with a smaller value in the inner region and a much
larger one in the outer disk (Chang et al. 1999; Boissier \&
Prantzos 1999; Hou et al. 2000; Chiappini et al. 2001). On the other
hand, many observations show that radial truncation exists in most
disk galaxies (e.g. Pohlen et al. 2002), which means there is a
change in the slope of total mass surface density and gas mass
surface density from a shallow exponential disk to a much steeper
one (de Grijs et al. 2001). This truncation radius $r_t$ appears to
evolve with the redshift $z$ of a galaxy (Azzollini et al. 2008),
which is a strong indication that the disk grows from smaller to
larger. For a simple phenomenological disk formation model, this
property could be described by a disk formation time scale $t_0$,
which is proportional to the radius from the galaxy center. In both
cases of $\tau(r)$ and $t_0$, there is a rapid increase of the metal
abundance during the early epochs in the inner disk, leading to a
steep abundance gradient. As time goes on, star formation
``migrates'' to the outer disk, producing metals and flattening the
abundance gradient there.

However, the mechanisms mentioned above could work together.
It is difficult to distinguish which one plays a more important role,
or if the scenario may be different for various disk galaxies. The
purpose of this paper is to understand which mechanism plays the
most important role in the origin and evolution of disk abundance
gradients in a galactic chemical evolution model. We concentrate
on the Milky Way disk because there are various observations on its
disk abundance, especially the recent PNe data (Maciel et al.
2003, 2005, 2006) and open clusters (Chen et al. 2003; Carraro et
al. 2007), which provide unique constraints on time evolution of
the disk abundance gradient and that could be sensitive to different
model assumptions. In section 2, we will describe the main content of
our model, which includes the mechanisms mentioned above. In section
3, the resulting model radial profiles will be given, and the effect of
$\tau$, SFR, and $t_0$ on the abundance gradient and its evolution will
be analyzed. The best model will be decided by comparing the model
results with the observational data. Finally, the main results will
be summarized in section 4.

\section{The Model}

We assume that the Milky Way has been embedded in a dark matter
halo. Primordial gas within the dark halo cools down to form a
rotationally supported disk, which is assumed to be sheet-like and a
system of a series of independent rings with the width of 500~pc.
The ring centered at $r_\odot=8$ kpc is labeled as the solar
neighborhood. Star formation and chemical evolution proceed
in each ring due to infalling gas.

\subsection{Infall Rate and Disk Formation Process}

The assumption of a gas infall process is traditionally based upon the
need to explain the locally observed metallicity distribution of
long-lived stars, which cannot be explained by the simple
``closed-box'' model (leading to the well-known ``G-dwarf problem'',
Pagel 1989). Recent observations show that the Milky Way disk (and
also M31, M33) is currently accreting substantial amounts of gas
with low metallicity ($\sim$ 1 \msyr) from high velocity clouds
(Blitz et al. 1999; Ferguson et al. 2002; Thilker et al. 2004; Ibata
et al. 2005; van den Bergh 2006). We note that Sancisi et al.(2008)
proposed that the accreting rate of the Milky Way could be about
0.1$\sim$ 0.2 \msyr, about ten times less than that required by the
observed SFR in the disk. However, there is some evidence
that the extra gas may come from a Galactic fountain (Sancisi et al.
2008).

Like many other phenomenological models, we also adopt an
exponentially decaying gas infall rate $f(r,t)$ as:

\begin{equation}\label{eq:infall}
f(r,t) =
\begin{cases}
A\left( r \right)e^{ - \left[ {t - t_0 \left( r \right)} \right]/\tau(r) }
& t \ge t_0 \left( r \right)  \\
0 & t < t_0 \left( r \right)
\end{cases}
\end{equation}
where $t_0(r)$ is the time when gas at radius $r$ begins to infall
onto the disk. $A(r)$ is normalized $\int_0^{t_g }
{f(r,t)dt}=\Sigma_{\rm{tot}}(r,t_g)$, where
$\Sigma_{\rm{tot}}(r,t_g)$ is the current total (gas+star) mass
surface density, and $t_g$ is the age of the disk. We adopt
$t_g=13.5$~Gyr for the Milky Way. That means the Milky Way Galaxy
formed at the beginning of the universe. We note that the thin disk
is relatively young on average, but the oldest open clusters in the
disk could be as old as 10 Gyr (Chen et al. 2003; Carraro et al.
2007), while dwarf stars in the disk have ages as old as that of
Galaxy.
As a working hypothesis, we shall always assume that the disk is as
old as the Galaxy (see also in Boissier and Prantzos 1999; Hou et
al. 2000). Test calculations show that adopting a smaller disk age
(for example, 11~Gyr) does not effect the main results of the model.
This is reasonable since the disk evolves very little in the last
several Gyr.

We assume that the current total surface density (unit: \mspc) of
the disk has an exponential form:
\begin{equation}\label{eq:sigma}
    \Sigma _{{\rm{tot}}} (r,t_g ) = \Sigma _{{\rm{tot}}}
    (r_ \odot  ,t_g )e^{ - (r - r_ \odot  )/r_d }
\end{equation}
From Equations (\ref{eq:infall}) and (\ref{eq:sigma}), we can get:
\begin{equation}\label{eq:sigmatot}
A(r) = \frac{{\Sigma _{\rm{tot}} (r_ \odot  ,t_g )e^{ - (r - r_
\odot )/r_d } }} {{\left[ {1 - e^{ - \left[ {t_g  - t_0 (r)}
\right]/\tau (r)} } \right]\tau \left( r \right)}}
\end{equation}
where $r_d$ is the disk radial scale length, and
$\Sigma_{\rm{tot}}(r_\odot,t_g)$ is the total mass surface density
in the solar neighborhood at the present-day. We adopt $r_d=2.7$kpc
and $\Sigma _{{\rm{tot}}} (r_ \odot  ,t_g ) = 56$\mspc\ for the
Milky Way disk (Robin et al. 1996; Boissier \& Prantzos 1999;
Holmberg \& Flynn 2004; Hammer et al. 2007). The combination of
Eq.(\ref{eq:infall}) and Eq.(\ref{eq:sigmatot}) will fix the history
of gas infall of disk if $t_0(r)$ and $\tau(r)$ are given.

Two cases in the above model correspond to the so-called
``inside-out'' disk formation scenario. One is represented by the
radial dependent infall time scale, e.g. $\tau(r) = a \times r+b$
with $a>0$, which means the infall time scale of the inner disk is
shorter than that of the outer disk. The second is described by
radial dependent disk formation time scale $t_0(r)$, which is
somewhat similar to the time dependent disk truncation. This case,
although over-simplified, is supported by both observations
(Azzollini et al. 2008) and numerical simulations (Ro\v{s}kar et al.
2008). The real mechanisms for the formation and evolution of the
disk truncation are quite complicated and not well understood (as
discussed in Azzollini et al. 2008; Franx et al. 2008; Ro\v{s}kar et
al. 2008). Here we just assume the disk size increases with time,
i.e. $t_0$ increases with galactic-centric distance $r/r_d$. Thus we
adopt a simple linear form:
\begin{equation}\label{eq:t0}
t_0(r)=\gamma \times(r/r_d)
\end{equation}
where $\gamma$ is a free parameter. This assumption corresponds to
the gradual build-up of the disk from the inner to the outer part.
Our purpose is try to understand how this disk truncation influences
the disk abundance gradient evolution. In Section 3, we shall
discuss how $t_0(r)$ and $\tau(r)$ affect the build-up of radial
abundance gradient.

\subsection{Star Formation Law}

Kennicutt (1998) has shown that the global SFRs of disks and
circumnuclear starburst galaxies are correlated with the local gas
density. The entire range of investigated galaxies, spanning a
magnitude of 5-6 orders in gas mass and SFR surface densities, fits
on a common power law with index $n \sim 1.4$. The tight relation
shows that a simple Schmidt (1959) power law provides an excellent
empirical parameterization of the SFR across an enormous range of
SFRs. It also suggests that the gas density is the primary
determinant of the SFR on these scales. Models of galaxy formation
and evolution usually adopt this best-fitting Kennicutt-Schmidt law
(hereafter, Classical KS law):
\begin{equation}\label{eq:KSLaw}
    \Psi(r,t)=0.25\Sigma_{\rm{gas}}^{n}(r,t)
\end{equation}
where $\Sigma_{\rm{gas}}$ is in unit of \mspc. $\Psi(r,t)$ is in
unit of \mspcGyr, $n=1.4\pm0.15$.

From the observational point of view, Kennicutt (1998) also found a
correlation between the observed SFR and the ratio of the surface
gas density to the local dynamical time scale, $\Sigma _{{\rm{gas}}}
/\tau _{{\rm{dyn}}} $. In particular, from this correlation one can
derive the following parametrization:
\begin{equation}
    \Psi (r,t) \propto \frac{{\Sigma _{{\rm{gas}}} }}
    {{\tau _{{\rm{dyn}}} }} \propto \Sigma _{{\rm{gas}}} \Omega
\end{equation}
where $\Omega$ is the angular rotation speed of the gas. Since
$\Omega \sim V(r)/r$, the SFR could be expressed as $\Psi (r,t)
\propto \Sigma _{{\rm{gas}}} \frac{{V(r)}}{r}$, where $V(r)$ is the
disk circular velocity at radius $r$.

In Boissier \& Prantzos (1999), the SFR was also expressed as
(hereafter, Modified KS law):
\begin{equation} \label{eq:MKSLaw}
\Psi(r,t) = \alpha \Sigma_{\rm{gas}}^{n}
\left[\frac{V(r)}{220}\right] \left[\frac{r}{r_\odot}\right]^{-1}
\end{equation}
where $\Sigma_{\rm{gas}}$ is in unit of \mspc, $V(r)$ is in unit of
\kms, $r$ is in unit of kpc. The index $n$ is chosen to be $\sim$
1.5 on an empirical basis. And for the Milky Way disk, $\alpha$ =
0.1, $V(r) = V_{{\rm{max}}} = 220$\kms. They also adopted this
Modified KS law in subsequent models for external spirals, and their
models can successfully reproduce the global properties of spirals
(Boissier \& Prantzos 2000, 2001; Boissier et al. 2001).


Concerning the average properties of SFR and gas mass surface
densities, the Classical KS law was further supported by recent
observations. Boissier et al. (2007) have investigated 43 nearby,
late-type galaxies from GALEX and conclude that, on the average,
their results are compatible with Classical KS law and can extend
this simple law to much lower gas mass surface density. But for an
individual galaxy, some will show a very untypical pattern in the
lg(SFR)-lg($\Sigma_{gas}$) plot, e.g. the M31 disk, as pointed out by
Boissier et al. (2007) and Yin et al. (2008).

In an earlier paper, Molla et al.\ (1997) also investigated the
radial abundance distributions for a number of nearby spirals,
including the Milky Way. In treating the star formation processes
for different galaxies, they considered the radially dependent
efficiency for the formation of molecular clouds. The efficiencies
are allowed to change from galaxy to galaxy. In this work, we have
simply adopted the classical KS law (Eq.(\ref{eq:KSLaw}), Kennicutt
1998) and modified KS law (Eq.(\ref{eq:MKSLaw})), where the later case is
equivalent to a radial dependent efficiency of star formation. This
kind of modified SFR has been extensively adopted during the past
for chemical evolution studies of the Milky Way disk (see review
by Matteucci 2001). But this is not to say that the classical K-S
law does not properly in describe disk galaxies. In fact, the
classical KS law has been widely accepted for the average star formation
properties of a galaxy. In many studies, especially in
semi-analytic modeling of disk galaxies, the classical K-S law is
often adopted. The recent space resolved observations for M51a by
Kennicutt et al. (2007) show that the index is comparable with the
classical KS law, but this does not mean it is valid for all
individual galaxies, especially in the past history of a galaxy. A
comparison study is needed based on the disk evolution history.
The time evolution of the abundance gradient and how it changes with
galactocentric distance are good constraints. These kind of constraints
have been used less often because there are fewer observational
constraints. Our purpose is try to understand which kind of SFR is
more suitable to describe the local star formation properties in
the Milky Way disk by calculating the evolution history of the disk
abundance gradient.

\begin{deluxetable}{ccc}
\tabletypesize{\small} \tablewidth{0pc} \tablecaption{Parameters
Adopted in the Models \label{tab:1}} \tablehead{\colhead{Parameters}
& \colhead{Values} & \colhead{References}} \startdata
$r_d$ (kpc) & 2.7 & Robin et al. (1996) \\
$R$ & 0.32 & Kroupa et al. (1993) \\
$12+\log\ ({\rm{O/H}})_ \odot$ & 8.7 & Lodders (2003)\\
$\Sigma _{{\rm{tot}}} (r_ \odot  ,t_g )$ (\mspc) & 56 & Holmberg \& Flynn (2004)  \\
$t_g$ (Gyr) & 13.5 &  WMAP3  \\
$y_i$ & $\left( {X_i } \right)_ \odot$ &   \\
\enddata
\end{deluxetable}

\begin{deluxetable}{cccl}
\tabletypesize{\small} \tablewidth{0pc} \tablecaption{SFR and timescales
for three models \label{tab:2}} \tablehead{\colhead{Model} &
\colhead{SFR} & \colhead{$t_0 (r) = \gamma \frac{r}{{r_d}}$} & \colhead{$\tau(r)$ (Gyr)}} \startdata
Mod-A  & Classical KS law & $ \gamma = 0   $ &  1.0,$\infty$, $0.75r+1$  \\
Mod-B  & Classical KS law & $ \gamma = 1.9 $ &  1.0,$\infty$, $0.75r+1$  \\
Mod-C  & Modified KS law  & $ \gamma = 0   $ &  1.0,$\infty$, $0.75r+1$  \\
\enddata
\end{deluxetable}

\subsection{Chemical Evolution}

The galactic disk is considered as an ensemble of concentric,
independently evolving rings, progressively built up by infall of
primordial composition. For the purpose of simplicity, we adopt the
Instantaneous Recycling Approximation (IRA). Therefore, the chemical
evolution at each ring can be followed by solving these appropriate
set of integro-differential equations (Prantzos 2008):
\begin{equation}\label{eq:GCE}
\left\{ {\begin{array}{*{20}c}
   {\frac{d}{{dt}}\Sigma _{{\rm{gas}}}  =  - \left( {1 - R} \right)\Psi  + f}  \\
   {\frac{d}{{dt}}\left( {X_i \Sigma _{{\rm{gas}}} } \right) =  - X_i \Psi  + E_i }  \\
\end{array}} \right.
\end{equation}
where $X_i$ is the mass abundance of element $i$; $R$ is the return
fraction; $E_i$ is the rate at which dying stars restore both the
enriched and unenriched material into the ISM. When IRA is adopted,
$E_i=y_i\Psi+(X_i-y_i)\Psi R$, where $y_i$ is the yield of element
$i$ and we adopt $y_i=\left( {X_i } \right)_ \odot$ throughout this
paper. Adopting z stellar Initial Mass Function (IMF) as a multi-slope
power-law between 0.1 \ms \ and 100 \ms \ from the work of Kroupa et
al. (1993), we can get $R$ = 0.32.

The metals predicted by the model are compared with observed
abundances of alpha elements (e.g. oxygen), which are mainly
produced by SN II explosions of massive stars. In this paper, when
we compare with the observed abundance gradients, we mainly refer to
oxygen abundance. We shall convert the abundance gradient of iron
from open clusters to that of oxygen according to the calibration of
Maciel et al. (2005).

\section{Model Parameters and Results}

The input parameters adopted in the model are listed in
Tab.\ref{tab:1}. In order to distinguish which mechanism plays the
most important role in the origin and evolution of abundance
gradients, we adopt 3 models for different SFRs and disk formation
time scale prescriptions which are listed in Tab.\ref{tab:2}. In
each Model, three forms of the infall time scale are considered. One is
$\tau=1.0$Gyr ($\tau=0$ is not adopted for it is the ``closed-box''
model which leads to G-dwarf problem) and another one is
$\tau=\infty$, both of which represent extreme cases for constant
infall time scale. The third one is ``inside-out'' scenario, we
adopt $\tau(r) = 0.75\times r+1 $ (Gyr), where $r$ is in unit of
kpc.

\subsection{Radial Profiles of Gas, Star and SFR}

\begin{figure*}[!t]
  \centering
  \includegraphics[height=16cm,width=10.5cm,angle=90]{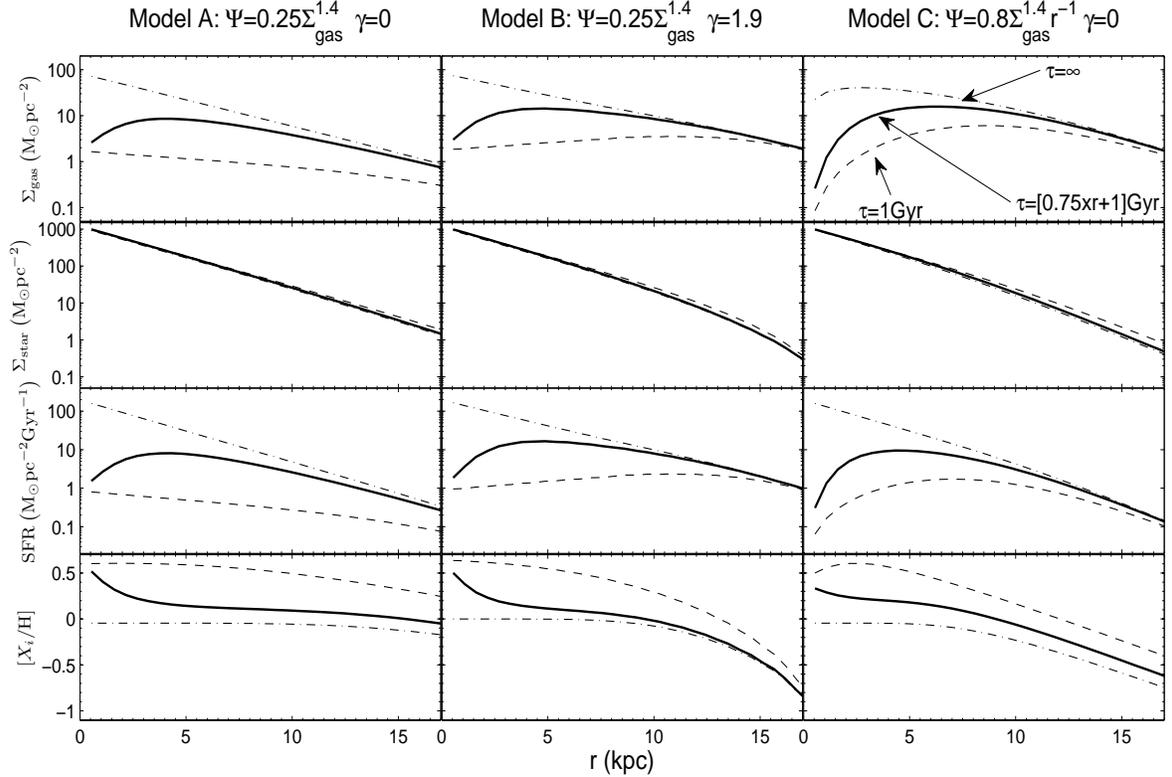}\\
\caption{The current radial profiles of gas mass surface density
$\Sigma_{\rm{gas}}$ (\mspc), stellar mass surface density
$\Sigma_{\rm{star}}$ (\mspc), star formation rate (\mspcGyr), and
abundance $\left[ {X_i /{\rm{H}}} \right]$ of element $i$. Three
columns represent 3 models mentioned above, which are labeled on the
top. Disk formation time scale is given by $t_0(r)=\gamma \times
r/r_d$. Infall time scale is given by $\tau(r) = 0.75\times r+1$
(Gyr) (solid line). In all panels, the dashed and dash-dotted lines
represent $\tau=1$ Gyr and $\tau=\infty$ respectively under constant
$\tau$ assumption. \label{fig:radial}}
\end{figure*}

The results for the present-day radial profiles of the 3 models are
plotted in Fig.\ref{fig:radial}.  These profiles include gas mass surface
density $\Sigma_{\rm{gas}}$, stellar mass surface density
$\Sigma_{\rm{star}}$, star formation rate and abundance $\left[ {X_i
/{\rm{H}}} \right]$ of element $i$. In all panels of
Fig.\ref{fig:radial}, the dashed and dash-dotted curves are two
limits of the constant $\tau$: dashed ones represent $\tau=1$Gyr and
dash-dotted ones represent $\tau=\infty$. Results of other constant
$\tau$ are between the two limits. The results of the ``inside-out''
scenario are plotted with solid lines.

From Fig.\ref{fig:radial}, we can see that a curve of radially
dependent $\tau(r)$ is actually the ensemble of points on curves of
different constant $\tau$, and each point at an arbitrary radius
$r'$ is the point at $r'$ on the curve of constant $\tau=\tau(r')$.
Thus no matter what form of $\tau(r)$ is adopted, the radial
profiles of all quantities from radius dependent $\tau(r)$ are
between the profiles of maximum and minimum constant $\tau$. From
Fig.\ref{fig:radial}, we can also find that the shape of a radial
profile is mainly determined by the adopted form of SFR and
$t_0(r)$, while different values and forms of $\tau(r)$ give similar
profiles. We also notice that Model C predicts a deep central
depression of the gas mass surface density as long as the infall
time scale is not too big. This is because the equivalent star
formation efficiency (that is $\alpha \times (1/r)$) is inversely
proportional to the distance from the galaxy center, therefore,
there is a large and fast consumption of infalling gas in the inner
disk.

\begin{deluxetable}{cccc}
\tabletypesize{\small} \tablewidth{0pc} \tablecaption{Model results
of current disk abundance gradient (unit: \dkp)\label{tab:3}}
\tablehead{\colhead{$\tau$ (Gyr)}&\colhead{Mod-A} & \colhead{Mod-B}
& \colhead{Mod-C}} \startdata \multicolumn{4}{c} {whole disk:
4.0-16.2 kpc}\\ \hline
$\tau=1$ & -0.027  & -0.091 & -0.075\\
$\tau=\infty$ & -0.009  & -0.056 & -0.054\\
0.75$\times r$ + 1 & -0.015  & -0.066 & -0.063\\
\hline \multicolumn{4}{c} {inner disk: 4.0-8.0 kpc}\\ \hline
$\tau=1$ & -0.015 & -0.040 & -0.065\\
$\tau=\infty$ & -0.001 & -0.007 & -0.023\\
0.75$\times r$ + 1 & -0.013 & -0.021 & -0.036 \\
\hline \multicolumn{4}{c} {outer disk: 8.0-16.2 kpc}\\ \hline
$\tau=1$ & -0.032 & -0.115 & -0.079 \\
$\tau=\infty$ & -0.013 & -0.079 & -0.068\\
0.75$\times r$ + 1 & -0.017 & -0.087 & -0.076\\
\enddata
\end{deluxetable}

\subsection{Radial Abundance Gradients}

We calculate the current abundance gradients $\frac{{d\left[ {X_i
/{\rm{H}}} \right]}}{{dr}}$ from the inner part, $r=4.0$ kpc, to the
outer part, $r=16.2$ kpc, of the disk. This range is defined as the
whole disk, which is chosen in order to be consistent with most of
the observations. We also divide the disk into inner and outer parts
with the intersection at solar position $r_{\odot}=8$kpc. The
results for all 3 models are listed in Tab.\ref{tab:3}. We shall
discuss the effect of $\tau(r)$, $t_0(r)$ and SFR on the abundance
gradients, respectively. Test calculations show that changing the
outer disk radius (for example, up to 21~kpc, about 8$r_d$) does not
affect the main conclusions discussed below.

{\bf Infall Time Scale, $\tau(r)$:} From Tab.\ref{tab:3}, we find
that in all cases, for a constant infall time scale, the abundance
gradients get flatter with the increase of $\tau$ value. When
$\tau(r)$ is radially dependent, results are between the two extreme
cases ($\tau$=1Gyr and $\infty$). In Mod-A, the current gradient
value is much smaller than other two cases (we will see below that
Mod-A is not consistent with the observed value in the Milky Way
disk). In Mod-B and C, the gradient values of the whole disk are
similar for different infall time scale. We will see below that
compared with the role of SFR and disk formation time scale
$t_0(r)$, infall time scale plays minor role in shaping the
abundance profile in the Milky Way disk.

{\bf Disk Formation Time Scale, $t_0(r)$:} Now, we turn to the disk
formation time scale $t_0$. In Mod-B, we adopt the same SFR form as
in Mod-A (Classical KS law), but we introduce the
disk formation time scale $t_0(r)$. From Fig.\ref{fig:radial} and
Tab.\ref{tab:3}, we find that the Classical KS law with an
appropriate $t_0(r)$ (hereafter we assume $t_0(r)=1.9 \times r/r_d$
Gyr) will produce an abundance gradient consistent with the
observation in the Milky Way disk ($- 0.05 \sim - 0.07$ \dkp). We
can also see that different infall time scales do not affect the
final gradient much.

{\bf Star Formation Rate:} In Mod-C, we have adopted the modified KS
star formation law, that is the radially dependent SFR. But we do not
introduce the disk formation time scale, i.e. $t_0(r)=0$. From
Fig.\ref{fig:radial} and Tab.\ref{tab:3}, we can see that this model
can also produce a strong enough abundance gradient for a disk evolved to
the present day.  Again, we find that infall time scales play a minor role.

In summary, we can see that when we adopt the classical KS star
formation law ($\Psi\ \propto\ \Sigma_{\rm{gas}}^{n}$), the chemical
evolution model based on the so-called ``inside-out'' infall time
scale cannot produce an abundance gradient that is steep enough at
the present day. If we introduce the disk formation time scale, then
the model is able to produce a steep abundance gradient no matter
how we choose the infall time scale. On the other hand, when we
adopt the radial dependent star formation law, i.e. the modified KS
law ($\Psi\ \propto\ \Sigma_{\rm{gas}}^{n}r^{-1}$), the model is
able to predict the required abundance gradient by choosing proper
infall time scale. From the results of three models, we can find
that when $t_0(r)\ne 0$, an appropriately steep abundance gradient
can be produced no matter how we choose the infall time scale $\tau$
and star formation law.

In Tab.\ref{tab:3}, we also present the abundance gradients for
inner and outer disk. We divide the disk into inner and outer part
at radius $r_\odot=8$ kpc. For all models, the outer disk has
larger abundance gradient than inner part, especially in Mod-B and
Mod-C. This is because in Model B, we have introduced the disk
formation time scale, which results in a delayed star formation and
lower metallicity in the outer disk. While in Mod-C, SFR is
inversely proportional to radial distance from the galactic center,
which results in lower a star formation rate and lower metal
production rate in the outer disk.

\begin{deluxetable}{ccccl}
\tabletypesize{\small} \tablewidth{0pc} \tablecaption{Observed
abundance gradients in the outer disk \label{tab:4}}
\tablehead{\colhead{Tracers} & \colhead{Element}   & \colhead{$R_g$}
& \colhead{Gradients} & Refs. \\
          &         &       (kpc)   & (dex/kpc) &  }
\startdata
Cepheids  &    Fe   &       8-12    &  -0.061   &    (1)     \\
Cepheids  &    O    &       8-12    &  -0.041   &    (1)     \\
Cepheids  &    Fe   &       8-12    &  -0.056   &    (1),(2) \\
Cepheids  &    O    &       8-12    &  -0.051   &    (1),(2) \\
Cepheids  &    Fe   &       10-16.4 &  -0.077   &    (2)     \\
Cepheids  &    Fe   &       10-15   &  -0.012   &    (3)     \\
Cepheids  &    Fe   &       10-15   &  -0.050   &    (2),(3) \\
Cepheids  &    Fe   &       12-17.2 &  -0.052   &    (4)  \\
Open Clusters & Fe  &       8-14    &  -0.06    &    (5)  \\
Open Clusters & Fe  &       12-22   &  -0.018   &    (6)  \\
Open Clusters & $\alpha$ elements & 12-22  & -0.003$\sim$-0.019 & (6) \\
Open Clusters & Fe  &       10-16   & -0.047    &    (7)  \\
\enddata \\
Ref: (1)Lemasle et al. 2007; (2)Andrievsky et al. 2002; (3)Lemasle
et al. 2008; (4)Yong et al. 2006; (5)Friel et al. 2002; (6)Carraro
et al.2007; (7)Chen et al. 2003
\end{deluxetable}

On the observational side, current findings on the shape of present
disk abundance gradient, especially in the outer part of the disk,
are still quite controversial (see Table.4 for a compilations from
the recent literature). Recent observations of the outer disk open
clusters have shown a flattening of the Galactic gradient beyond
10-12kpc (Carraro et al. 2007; Yong et al. 2005). However, other studies
also based on the open clusters show that the gradient of
outer disk is steeper (Janes et al. 1988; Friel et al. 2002; Chen et
al. 2003). Other studies based on different tracers, such as HII
regions (Vilchez \& Esteban 1996), Cepheids (Andrievsky et al. 2004;
Lemasle et al. 2008), planetary nebulae (Costa et al. 2004), also
indicate some hints of flattening gradient in the outer disk. On the
other hand, there are also studies based on young stars and HII
regions showing a continuous decreasing of gradient toward the outer
disk (Deharveng et al. 2000; Rolleston et al. 2000).

When compared with our model predictions, an observed small gradient
in the outer disk is compatible with the value from Mod-A, while a
steeper gradient is consistent with Mod-C. Clearly, much work needs
to be done on both observations and modeling. The time variation of the
Galactic abundance gradient could provide a more comprehensive test
for the different chemical evolution models than just a single
measured abundance gradient at the current time. Also abundance
measurements of $\alpha$ elements for outer disk tracers are very much
needed in order to
have detailed abundance ratio analysis, such as the radial variation of
[O/Fe]. This is very important to understand the history of star
formation in the outer disk since $\alpha$ elements are mainly
produced in massive stars when they exploded as SNII, while iron is
mainly synthesized from low and intermediate stars exploded as SNIa.
Those two kinds of elements may have different gradient behavior.
Cepheids are high-mass stars whose abundance values could be
regarded as the current composition of the interstellar medium.
Although a number of abundance measurements have been done by
observing Cepheids in the outer disk, the current situation is still
not very clear. For example, the outer disk Cepheids appear to
exhibit a bimodal distribution for the [Fe/H] and [O/Fe] (Yong et
al. 2006; Lemasle et al. 2008). The single linear radial abundance
gradient is also questioned by observations from Cepheids
(Andrievsky et al. 2004; Lemasle et al. 2008) and open clusters
(Twarog et al. 1997). There are also alternative points of view
regarding the formation history of the outer Galactic disk, in which
merger events or accretions of small satellites may be responsible
for the unusual metallicity distributions observed in outer open
clusters and Cepheids (Yong et al. 2005). If this is the case, much
sophisticated infall models are needed.

\begin{figure*}[!t]
  \centering
  \includegraphics[angle=90,height=10cm,width=14cm]{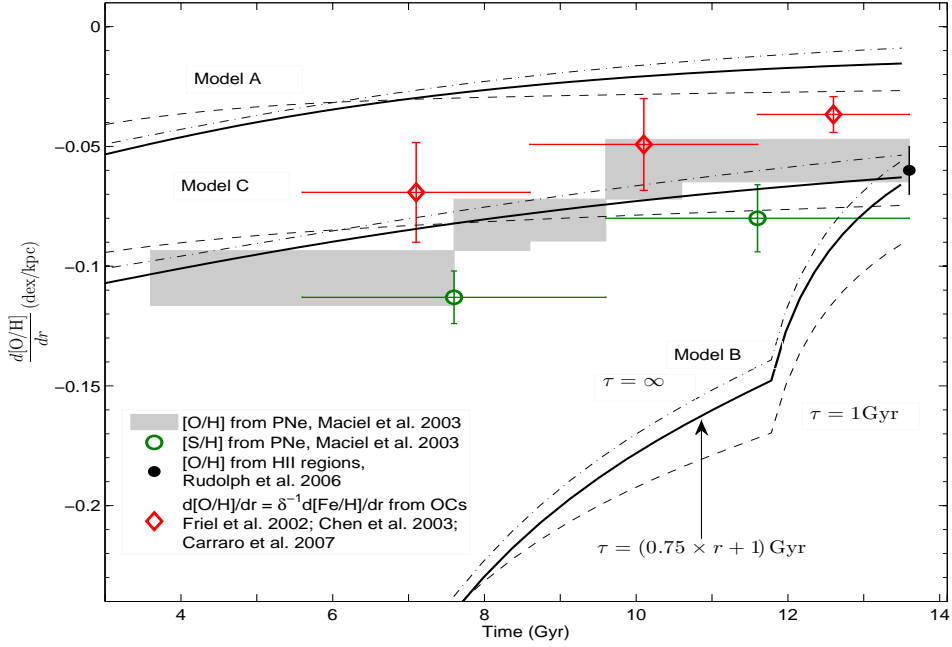}\\
\caption{Time evolution of the radial abundance gradient.
Predictions of three models (Model A, B and C) are plotted. Dashed,
dash-dotted and solid lines represent different infall time scale of
$\tau=1$Gyr, $\tau=\infty$ and $\tau(r) = (0.75\times r+1)$ Gyr
respectively. Three models show quite different history of abundance
gradient. The disrupted rise in Model B at about 11.8 Gyr is caused
by the disk formation time scale. The shaded area is the observed
oxygen abundance gradients from PNe (Maciel et al. 2003). Open
circles are sulphur abundance gradients from different aged PNe
(Maciel et al. 2003). The open diamonds are oxygen abundance
gradients from open clusters samples between $r=4.0$ kpc to $r=16.2$
kpc, which are transferred from iron abundance gradients according
to the calibration of Maciel et al. (2005). In order to overcome the
shortage of outer disk cluster samples in Chen et al. (2003), we
have combined the data from Friel et al. (2002), Chen et al. (2003)
and Carraro et al. (2007). The black point is the current oxygen
abundance gradient obtained from HII regions (Rudolph et al. 2006).
\label{fig:time}}
\end{figure*}

\subsection{The Evolution of Abundance Gradient}

As we have already mentioned above, Mod-B and C can both produce
the proper current abundance gradient, which is consistent with the
observed values along the Galactic disk. It is difficult to conclude
which one is better.

Now we discuss the time evolution of the abundance gradient. For Mod-A
and C, we calculate the gradients $\frac{{d\left[ {X_i /{\rm{H}}}
\right]}}{{dr}}$ still from the inner part to the outer part of the
disk (4.0 - 16.2~kpc) like what we do with current gradient. But for
Mod-B, we calculate from the inner part (4.0~kpc) to the outermost
boundary where the disk just begins to form by infall gas at time
$t=t_0$. When the disk grows up to 16.2~kpc, we fix the radius
range to be between 4.0 and 16.2~kpc. This is also why the curves of
Model B in Fig.\ref{fig:time} show an abrupt increase in the
gradient around 11.8 Gyr, since the disk has not grown to 16.2 kpc
before this stage.

The predicted evolutions of the abundance gradients for three models
are plotted in Fig.\ref{fig:time}. For each model, we calculated
three cases for infall time scale, as was indicated in the figure.
The shaded area is the observed oxygen abundance gradient from
different PN populations (Maciel et al. 2003; 2006). In Chen et al.
(2003), a shallow current iron abundance gradient is derived. In checking
their data, however, we find that the gradient value
heavily relies on one outer young cluster at about 15~kpc. There are
not enough young clusters in the sample at large radius. In order to
overcome this shortage, we have combined the data of open clusters
from Friel et al. (2002), Chen et al. (2003) (mainly the OCs of
$r<13$kpc) and Carraro et al. (2007) (mainly the OCs of $r>12$kpc),
then we show the evolution of gradients obtained from the samples of
4.0-16.2 kpc, where the iron gradient has been transferred to oxygen
by the following calibration (Maciel et al. 2005):
\begin{equation}\label{eq:FetoO}
\frac{{d\left[ {12 + \log \left( {{\rm{O/H}}} \right)} \right]}}{{dr}}
= \frac{1}{\delta }\frac{{d[{\rm{Fe/H}}]}}{{dr}}
\end{equation}
where $\delta$ is the coefficient independent of the galactocentric
distance, and is about 1.0$\sim 1.5$. Here we adopt $\delta$=1.2
(Maciel et al. 2005).

Based on Fig. \ref{fig:time}, we can find that:

1) All three models show that the disk abundance gradient is steeper
in the early stage of disk evolution. Both Model B and C can predict
acceptable current abundance gradients when they are compared with
the young tracers in the Galactic disk.

2) Model A always predicts a flatter abundance gradient compared
with other models at any time. It also predicts a much shallower
current gradient compared with the results from young objects (e.g.
HII regions from Rudolf et al. 2006; Cepheids from Yong et al.
2006). We note that there are also a number of observations showing
a shallow radial abundance gradient (Deharveng et al. 2000; Daflon
\& Cunha 2004; Carraro et al. 2007), but the absolute values for
oxygen and iron are still larger than the predictions of Model A. So
for the Milky Way disk, if the classical KS law of star formation is
adopted, models are not able to reproduce enough of a disk abundance
gradient, even if the infall time scale is radially dependent.

3) Although Model B can produce a reasonable value of the current gradient by
properly adjusting the infall time scale, it predicts a much steeper
gradient in the past. This is not supported by current available
observations from planetary nebulae and open clusters. The steep
abundance gradient at early times is due to the fact that the disk
formation time scale in Model B is radial dependent which results in
a delayed gas infall process and star formation in the outer disk.

4) Model C can predict not only the current values of
abundance gradient along the Milky Way disk, but also the evolution
of gradient which is fairly consistent with observations from PNe
and OCs. Since the classical KS law was derived from the average SFR
and gas mass surface densities for a whole galaxy (Kennicutt 1998),
we expect that the modified KS star formation law, that is the
radial dependent SFR, would be more suitable when we deal with the
local star formation processes within a galaxy.

\subsection{Comparison with Other Observations}

\begin{figure*}[!t]
  \centering
  \includegraphics[angle=90,height=10cm,width=14cm]{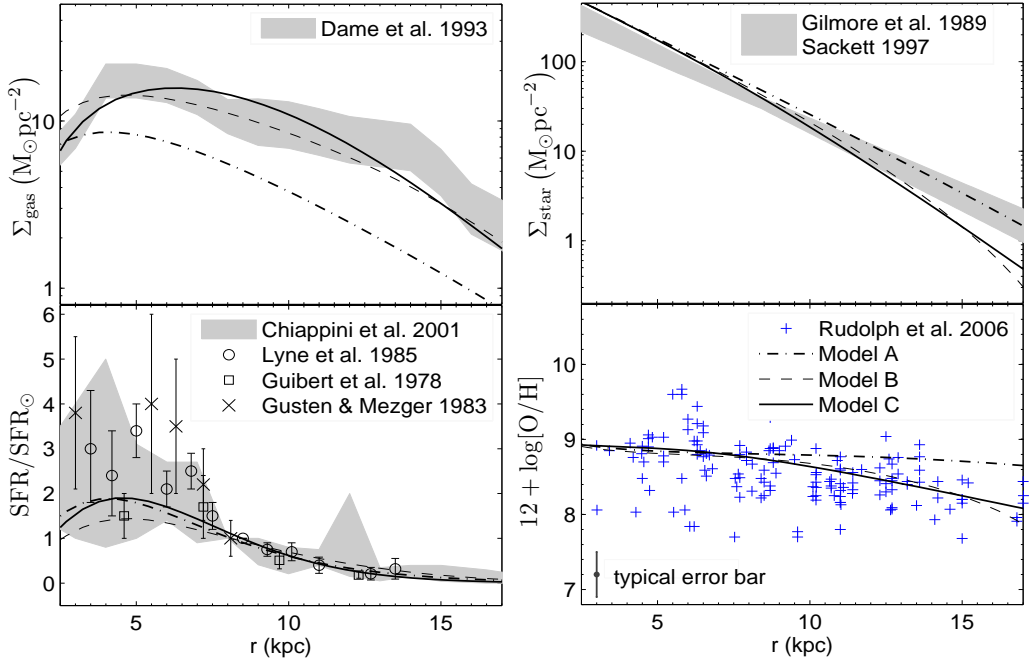}\\
\caption{Comparison of Model A, B and C with the observed profiles
of gas mass surface density, stellar mass surface density, star
formation rate and oxygen abundance of the Milky Way disk (from top
left to bottom right panel). The stellar mass surface density is
obtained by assuming an exponential disk profile, where the stellar
disk scale length $(r_d )_*=2.5-3$ kpc (Sackett 1997) and the
stellar mass surface density at the solar neighborhood $\left(
{\Sigma _{{\rm{star}}} } \right)_ \odot = 35 \pm 5$ \mspc\ (Gilmore
et al. 1989). For all three models, we adopt the infall time scale
to be $\tau(r) = (0.75\times r+1)$Gyr. \label{fig:compare}}
\end{figure*}

Now we further compare the model results with other constraints,
such as the gas, SFR and stellar profiles in the Milky Way disk.
According to the discussions given above, the form and the value of
infall time scale $\tau$ does not affect the radial profiles
greatly, so we shall adopt $\tau(r) = 0.75\times r + 1$. The results
for all three models are presented in Fig. \ref{fig:compare}.

The observational data of current gas surface density
$\Sigma_{\rm{gas}}$ come from Dame et al. (1993), ${\rm{SFR/SFR}}_
\odot$ from Gusten \& Mezger (1983), Lyne et al. (1985), Guibert et
al. (1978), and Chiappini et al. (2001). The data of
$\Sigma_{\rm{star}}$ are from the solar neighbourhood value $\left(
{\Sigma _{{\rm{star}}} } \right)_ \odot   = 35 \pm 5$ \mspc\
(Gilmore et al. 1989) and extend to the whole disk with an
exponential radial profile of stars, in which $(r_d )_ *=2.5 \sim 3$
kpc (Sackett 1997, Hammer et al. 2007). The oxygen abundance profile
is from Rudolph et al. (2006).

From Fig. \ref{fig:compare}, we can see that, in general, all of the
three models are able to predict a reasonable profile for
$\Sigma_{\rm{star}}$ and ${\rm{SFR/SFR}}_ \odot$. Since the observed
SFR and stellar mass surface density are quite uncertain, they are
not good tracers to constrain model results. However, based on the
observed gas and abundance profiles, it is clear that Mod-A is not a
good choice since it is difficult to get acceptable matches to the data no
matter how we modulate infall time scale. But both model B and C are
acceptable, and Model C is relatively much better, for it is
consistent with the above discussions based on the past history of
abundance gradient.

\section{Summary}

Based on a simple model of galactic chemical evolution, which
includes infall, star formation and delayed disk formation, we have
investigated the origin and evolution of the radial abundance
gradient of a disk galaxy. By comparing the model results with the
observations in the Milky Way disk, we try to find the effect of
each mechanism on shaping abundance gradients and decide which one
plays the main role.

The main results of our model can be summarized as follows:

\begin{enumerate}

\item For all adopted models and parameters, the predicted
      radial abundance gradients are steeper in the early epoch of disk
      evolution, which is consistent with observations of planetary nebulae and
      open clusters.

\item The disk formation time scale $t_0(r)$ could be the most important
      parameter governing the production of the abundance gradient, no matter what
      SFR and infall time scale are assumed. In this case, the inner disk accretes
      infalling gas earlier than the outer part,
      thus star formation and metal enrichment start earlier in the inner region.
      However, this will predict a very steep abundance gradient in the past, which
      is not consistent with the observations.

\item Besides $t_0(r)$, the radial dependence of SFR will
      further strengthen the gradient. If $t_0(r)$ is not considered
      (i.e. $\gamma=0$ in Eq.(\ref{eq:t0})), SFR plays the main role in
      shaping the disk abundance gradient. We find that the radially dependent
      SFR $ \Psi \propto \Sigma_{\rm{gas}}^{1.4} r^{-1} $ is a suitable
      model to fit the observational data in the Milky Way disk,
      while the classical KS law $ \Psi \propto \Sigma_{\rm{gas}}^{1.4}$
      without $t_0$ cannot produce a current abundance gradient
      steep enough to match those that have been observed in the Galactic
      disk from Cepheids and HII regions.

\item Relative to $t_0$ and SFR, the infall time scale $\tau$ only plays
      a minor role in shaping the abundance gradient. Given a
      certain SFR, the adopted form of $\tau(r)$ will not significantly
      change the gradient. But compared with constant infall time scale,
      an ``inside-out'' assumption for the time scale could help to produce some
      degree of abundance gradient.

\item When we divide the disk into inner and outer parts at the
      intersection radius of 8~kpc, our model predicts a steeper gradient in the
      outer disk. We note that there are observations showing a flatter
      abundance gradient for the outer Galactic disk based on
      different tracers, such as open clusters, Cepheids, and young
      stars. But there are also other independent studies that do not show
      evidence of a flattening gradient toward outer disk. Observations from Cepheids
      also show that the Galactic iron gradient might be more accurately described by a
      bimodal distribution, especially in the outer disk. All of those shows a
      complicated formation history of the outer Galactic disk. Further observational
      work and more sophisticated models need to be done before we can have a more
      clear understanding of the outer disk formation.

\end{enumerate}

In summary, we expect that for an individual Milky Way type galaxy,
a better description of the SFR is the modified KS star formation
law. In fact, the modified KS law can also be written as $ \Psi =
\alpha(r) \Sigma_{\rm{gas}}^{1.4}$. This is equivalent to classical
KS law, only the normalization coefficient $\alpha(r)$ (which is
related to star formation efficiency) is inversely proportional to
the distance from the galaxy center. This kind of SFR could well
reproduce the observed Galactic disk abundance gradient and its time
evolution.

\acknowledgements We are grateful to the critical comments from an
anonymous referee. Eric Peng is thanked for his helpful comments on
the writing of the paper. This work is supported by the National
Science Foundation of China No.10573028, 10573022, the Key Project
No.10833005, the Group Innovation Project No.10821302, and by 973
program No. 2007CB815402.

\def\apj{ApJ}
\def\apjl{ApJL}
\def\apjs{ApJS}
\def\aj{AJ}
\def\aap{A\&A}
\def\araa{ARA\&A}
\def\aapss{A\&AS}
\def\mnras{MNRAS}
\def\nature{Nature}
\def\apss{Ap\&SS}
\def\pasp{PASP}

{}

\end{document}